\documentclass[a4paper,12pt]{article}
\usepackage{amsfonts,amsthm,amsmath,amssymb}
\usepackage{graphicx}
\usepackage{braket}
\usepackage{bm}
\usepackage[subrefformat=parens]{subcaption}
\usepackage{mathrsfs}
\usepackage{booktabs}
\numberwithin{equation}{section}
\usepackage[
bookmarks=true,bookmarksnumbered=true,%
linkcolor=blue,anchorcolor=blue,urlcolor=blue,%
]{hyperref}

\newcommand{\be}{\begin{equation}}
\newcommand{\ee}{\end{equation}}
\newcommand{\bea}{\begin{eqnarray}}
\newcommand{\eea}{\end{eqnarray}}

\newcommand{\Tr}{\text{Tr}}

\newcommand{\ba}{\begin{eqnarray}}
\newcommand{\ea}{\end{eqnarray}}

\newcommand{\mcl}{\mathcal}

\newcommand{\f}{\frac}
\newcommand{\s}{\sqrt}

 \def\prt{\partial}

 \def\f {\frac}

\newcommand{\SU}{\mathrm{SU}}
\newcommand{\SO}{\mathrm{SO}}
\newcommand{\Spin}{\mathrm{Spin}}
\newcommand{\Sp}{\mathrm{USp}}
\newcommand{\SL}{\mathrm{SL}}
\newcommand{\PSL}{\mathrm{PSL}}

\newcommand{\cardy}[1]{|#1\rangle_{\mathsf{c}}}
\newcommand{\ishibashi}[1]{|#1\rangle\!\rangle}

\newcommand{\id}{\bm{1}}

\begin{document}

\begin{titlepage}
\thispagestyle{empty}

\begin{flushright}
OU-HET 956

\end{flushright}


\begin{center}
\noindent{{\textbf{Mixed global anomalies and boundary conformal field theories}}}\\
\vspace{2cm}
Tokiro Numasawa $^{1,2}$ and Satoshi Yamaguchi$^{1}$\vspace{1cm}

{\it
$^1$ 
Department of Physics, Graduate School of Science,\\
Osaka university, Toyonaka 560-0043, Japan
}
\vskip 1em
{\it 
$^2$
Department of Physics, McGill University,\\ 
3600 rue University, Montr\'eal, Qu\'ebec, Canada H3A 2T8}

\vskip 2em
\end{center}

\begin{abstract}
We consider the relation between mixed global gauge gravitational anomalies and boundary conformal field theory in WZW models for simple Lie groups.
The discrete symmetries of consideration are the centers of the simple Lie groups.
These mixed anomalies prevent gauging them i.e, taking the orbifold by the center.
The absence of anomalies impose conditions on the levels of WZW models.
Next, we study the conformal boundary conditions for the original theories.
We consider 
the 
existence of a conformal boundary state invariant under the action of the center. 
This also gives conditions on the levels of WZW models.
By considering the combined action of the center and charge conjugation on boundary states, we reproduce the condition obtained in the orbifold analysis.

\end{abstract}

\end{titlepage}
\newpage
\tableofcontents

\section{Introduction}

A 't Hooft anomaly is an obstruction to gauge a global symmetry, 
and puts a constraint on the RG flows 
called 't Hooft anomaly matching condition\cite{tHooft}; 
the 't Hooft anomaly of the IR theory matches with that of the UV theory 
as long as the theory has the global symmetry in question during the RG flow. 
This matching condition can also be applied to discrete symmetries\cite{CsMu}.

Recently, quantum anomalies are focus of attention in condensed matter physics, because they provide a useful tool to investigate symmetry protected topological (SPT) phases\cite{CGLW}.
When we put a theory in a non-trivial SPT phase on a manifold with boundaries, the boundary localized modes with 't Hooft anomalies appear.
They are now coupled to the bulk theory and 't Hooft anomalies are canceled in the same manner as the usual anomaly inflow mechanism\cite{CaHa}.

The study of SPT phases gives a natural motivation to consider the anomalies of discrete symmetries and global anomalies.
In the cases of perturbative anomalies we have them in the system with chiral fields in even dimensions.
On the other hand, anomalies of discrete symmetries and global anomalies can also exist in non-chiral, bosonic systems in odd dimensions.
Recently the anomalies for discrete symmetries are focus of attention\cite{KaTh1}\cite{KaTh2}, generalized to $p$-form symmetries\cite{GKSW}\cite{ThKe}, and applied to the study of non-supersymmetric gauge theories\cite{GKKS}\cite{TaKi}\cite{ShYo}.
In this paper, we consider the mixed anomalies between large diffeomorphisms and discrete symmetries.
The continuum counterpart of this mixed anomaly does not exist in $1+1$ dimensions because such mixed term can not appear in anomaly polynomials%
\footnote{Since a nontrivial Pontryagin class $p_k$ has degree 4 and we can not find such mixed term of $p_k$ and a Chern class $c_l$ in anomaly polynomials in degree 4 which is relevant for a 't Hooft anomaly in 1+1 dimensions.}.

Another application of a 't Hooft anomaly in condensed matter physics is to use itself as a classification of a gapless version of an SPT phase\cite{FuOs}.
If two theories have different 't Hooft anomalies for a symmetry $\Gamma$, they cannot be connected by RG flows while preserving the symmetry $\Gamma$ and thus they are in different symmetry protected phases.
In this manner, we can use the 't Hooft anomalies to detect both $(2+1)$-dimensional SPT phases\cite{CGLW} and $(1+1)$-dimensional symmetry protected critical phases\cite{FuOs}%
\footnote{There are some gapped theory with 't Hooft anomalies for discrete symmetry, the theory may be flow to non-trivial topological ordered states\cite{SeWi}\cite{TaYo1}\cite{TaYo2}.
There is also a possibility to flow to a theory where $\Gamma$ is spontaneously broken.
In these cases 't Hooft anomalies still prevent to flow to trivial gapped states.}.

In conformal field theory, the procedure to gauge a discrete global symmetry $\Gamma$ is known as the orbifold construction.
In this procedure, we exclude the non-invariant states under $\Gamma$ action (i.e. project the spectrum onto the invariant states) 
and also need to include the twisted sector (soliton sector) to preserve the modular invariance, 
which is the invariance under a class of large diffeomorphisms.
Once we determine the way to project onto the invariant states, the modular invariance determines the twisted sector and the projection operation on it.
Sometimes the twisted sector determined from modular invariance is not compatible with the action of  the symmetry $\Gamma$ and causes inconsistency.
This inconsistency is the mixed anomaly between large diffeomorphisms (modular transformations $\SL (2,\mathbb{Z})$) and the symmetry $\Gamma$
.


The related constraints to the modular invariance are the consistency of boundary conformal field theories\cite{Cardy}.
There are several motivation to consider the relation between 't Hooft anomalies and boundary states.

First, one way to distinguish SPT phases is putting a theory on a non-trivial background\cite{GKKS}, and another way is putting the theory on a manifold with boundary\cite{CGLW,AKLT1,AKLT2}.
For example, we can find that the spin $1$ Haldane chain, whose phase is realized in the Affleck-Kennedy-Lieb-Tasaki (AKLT) model\cite{AKLT1,AKLT2}, is in a non-trivial phase by putting the theory on a non-trivial monopole background (non-trivial Stiefel-Whitney class)\cite{GKKS} or putting the theory on a manifold with a boundary\cite{Kennedy}\cite{CGLW}.
Now, we can think of an 't Hooft anomaly itself as a classification of a gapless version of a SPT phase\cite{FuOs}.
Therefore, it should be useful to consider the analog of them in CFTs.
The analog of the former is putting a theory on a non-trivial gauge-gravitational background.
The analog of the latter is to consider the CFTs with boundaries.
The natural boundary conditions are conformal boundary conditions, which keep the half of full conformal symmetry.

Second motivation is that in some cases we can detect $(2+1)$-dimensional SPT phases ($(1+1)$-dimensional 't Hooft anomalies) from boundary states \cite{HTHR,Bultinck:2017iff}.
According to \cite{HTHR,Bultinck:2017iff}, if we find 
Cardy states invariant under symmetry transformations, 
the symmetry is anomaly free and they do not corresponds to the edge theory of non-trivial SPT phases.
On the other hand, if we cannot construct such boundary states, 
the symmetry has 't Hooft anomalies.
This condition seems to be closely related to construction of the twisted sectors.

These consideration brings us to study the symmetry property of boundary conformal field theories.
In this paper we consider the 't Hooft anomaly in WZW models for Lie groups $G$ whose center $\Gamma$ is a cyclic group.
We consider the anomaly of the center $\Gamma$.
Here we give a brief summary of this paper.
Depending on the level $k$ and groups $G$, the orbifold theory by $\Gamma$ may not be compatible with modular invariance, i.e. the large diffeomorphisms on a torus.
This can be understood as a mixed anomaly between the large differmorphisms and the center $\Gamma$.
Next, we study the existence of invariant boundary conditions under the center $\Gamma$ in the original WZW models with diagonal modular invariants for simple Lie groups $G$.
This also depends on the level $k$ and sometimes there are no such boundary states.
Surprisingly, these conditions obtained in the boundary state analysis perfectly match with the conditions obtained in the orbifold analysis for the simple group $G$ that does not have complex representations.
For simple Lie groups $G$ with complex representations $(A_n, D_{2l+1} ,E_6 )$, we consider the conditions for the existence of invariant boundary states under combined action of the generator $h \in \Gamma$ and charge conjugation $C$.
These conditions match with the conditions obtained in the orbifold analysis for these groups.

The rest of this paper is organized as follows.
In Section \ref{sec:orbifold}, we summarize how mixed global gauge gravitational anomalies appear in $(1+1)$-dimensional CFTs.
Only when this mixed anomalies are absent, we can take the orbifold.
Then, we review the case of the WZW model of Lie group $G$.
We consider the orbifold by the center and its subgroup.
We also show some examples of anomaly cancellation between two WZW models.
In section \ref{sec:BCFT}, 
we study the condition to find a symmetry invariant boundary state.
We show that we can find the same condition with the orbifold analysis.

\section{Mixed global anomalies and orbifold constructions in CFTs}\label{sec:orbifold}

\subsection{Coupling to external discrete gauge fields and mixed anomalies}
In this section, we consider the mixed gauge gravitational anomalies in $(1+1)$-dimensional CFTs.
This is nothing but the condition for the consistency of orbifold
\cite{FrVa}\cite{FGK}\cite{SCR}.
We review these results, emphasizing the point that mixed global gauge gravitational anomalies appear.

Let us consider a cyclic symmetry $\Gamma$ of order $N$ and its orbifold.
We consider the theory on a torus with modulus $\tau$.
First, we consider coupling the original theory to ``external gauge field,'' which means nontrivial twisted boundary conditions by $\Gamma$.
Because we consider CFTs on a torus, we can consider a twisted boundary condition on each direction.
We denote this boundary condition by $(h_t,h_x)$ where $h_t,h_x \in G$ denote the group elements to put the twisted boundary condition in imaginary time direction $t$ and space direction $x$ respectively.
In other words, the boundary condition for the fields $g(z,\bar{z})$ are given by
\ba
g(z+1,\bar{z}+1)  &=& h_x g(z,\bar{z}), \notag \\
g(z+\tau,\bar{z}+\bar{\tau})  &=& h_t g(z,\bar{z}).
\ea 

Partition functions with these twists are denoted as $Z_{(h_t,h_x)}$.
In operator formalism, $Z_{(h,\bf{1})}$ corresponds to the partition function with symmetry action $\Tr (h \,e^{-2\pi \text{Im}\tau H} e^{i 2\pi \text{Re}\tau P})$.
On the other hand, $Z_{(\id,h)}$ corresponds to the twisted sector partition function without the projection onto the gauge invariant states. 

Next we consider
modular transformations, which are the large diffeomorphisms on a torus.
Modular transformations $\SL (2,\mathbb{Z})$ are generated by $S = \begin{pmatrix} 0 & -1 \\ 1 & 0\end{pmatrix}$ and $T = \begin{pmatrix} 1 & 1 \\ 1 & 0\end{pmatrix}$.
We denote the action of $T,S \in \SL (2,\mathbb{Z})$ on a partition function $Z$ by $\mathcal{T}Z$ and $\mathcal{S}Z$.
The boundary condition $(h_t,h_x)$ is mapped to $(h_t^a h_x^b,h_t^ch_x^d)$ by
the modular transformation $\tau \to \frac{a \tau + b }{c \tau + d}$ labeled by $M = \begin{pmatrix} a & b \\ c & d\end{pmatrix} \in \SL (2,\mathbb{Z})$ 
\footnote{Though the action on modulus $\tau$ is projective and given by $\PSL (2,\mathbb{Z})$, 
the action on spacetime $T^2$ is given by $\SL (2,\mathbb{Z})$. 
Correspondingly, the map of boundary conditions depends on the sign of $M$ because $-\mathbf{1}$ flips the imaginary time and space.}. See for example\cite{FuOs}\cite{FrVa}.

If we assume this $\SL (2,\mathbb{Z})$ large diffeomorphism invariance, 
 partition functions $Z_{(h_t,h_x)}$ are related to each other by modular transformations 
according to the map of boundary conditions $(h_t,h_x) \to (h_t^a h_x^b,h_t^ch_x^d)$. 
Especially, this means $Z_{(\id,h)} = \mathcal{S} Z_{(h,\id )}$ and $Z_{(h^l,h)} = \mathcal{T}^l\mcl{S}Z_{(h,\id )}$.
The first equation means that the twisted sector partition function $Z_{(\id ,h)}$ is determined by symmetry action $\Tr (h\, e^{-2\pi \text{Im}\tau H} e^{i 2\pi \text{Re}\tau P})$ and its modular $S$ transformation.
The second equation means that the action of $\Gamma$ on the twisted sector is determined by the modular $T$ transformation.

The orbifold partition function is given by summing over the ``external gauge background'' configurations.
The summation over independent discrete gauge background is given by \footnote{Because we only consider the cyclic group $\mathbb{Z}_N$, we do not consider the existence of discrete torsion\cite{Vafa1986}\cite{BCR} .}
\be
Z_{orb} = \frac{1}{|\Gamma|} \sum_{h_t,h_x \in \Gamma} Z_{(h_t,h_x)}, \label{eq:orbifold}
\ee
where $|\Gamma|$ is the order of $\Gamma$ and corresponds to the gauge volume.

Until now we assumed that there are no anomalies.
Let us see how the above procedure fails when mixed anomalies exist.
An inconsistency can happens from the condition $Z_{(h^l,h)} = \mathcal{T}^l\mcl{S}Z_{(h,\id )}$.
Especially, we will see that in some cases $\mathcal{T}^N\mcl{S}Z_{(h,\id )} = Z_{(h^N,h)} =Z_{(\id ,h)} $ is not satisfied but pick a phase factor $e^{i\theta}Z_{(\id ,h)}$.
This means that the action of $h \in \Gamma$ on twisted sector determined from modular transformation actually  does not satisfy $h^N = 1$.
This non-trivial phase factor breaks the invariance under the transformation $ g(z,\bar{z}) \to h^{N}g(z,\bar{z})$, which is actually the identity transformation \cite{SCR}.
Therefore, the orbifold construction (\ref{eq:orbifold}) does not work in this case.

\subsection{$\SU (2)_k$ WZW models\label{SU(2)section}}
Here we consider the mixed global anomalies in $\SU (2)_k$ WZW models considered in \cite{FuOs}\cite{GeWi}.
$\SU (2)_k$ WZW models have $k+1$ primary fields which are labeled by the spin $0 \le j \le \frac{k}{2}$\footnote{There is also the label of the magnetic quantum number, but we do not need it and we omit it for simplicity.}.
The Lagrangian of this theory is given by\cite{Yellow} 
\be
S = \frac{|k|}{8\pi}\int dt dx \ \Tr[ \prt _{\mu} g \prt^{\mu} g^{-1 }] + \frac{k}{12\pi} \int_{M_3} \Tr [(\tilde{ g}^{-1}d \tilde{g})^3], \label{eq:LagWZWsu2}
\ee
where $g(t,x)$ takes the value on the $\SU (2)$ group manifold and $M_3$ in the second term is a $3$-dimensional manifold whose boundary is the spacetime we are considering.
$\tilde{g}$ is an extension of $g$ to $M_3$.
In order to be independent from the choice of $M_3$ and $\tilde{g}$, $k$ must be quantized correctly i.e. $k \in \mathbb{Z}$.
From now we only consider non-negative $k$.

In the diagonal theory, we have $k+1$ primary states $\ket{j,j}$ for $0 \le j \le \frac{k}{2}$.
The action of the center $\Gamma = \mathbb{Z}_2 = \{\id ,h\}$ is given by $h \ket{j,j} = (-1)^{2j}\ket{j,j}$.
The partition function of the diagonal theory is given by 
\be
Z_{diag} = Z_{(\id ,\id )} = \sum_{j \in \frac{1}{2}\mathbb{Z}} ^{\frac{k}{2}} |\chi_j|^2,
\ee
where $\chi_j$ is the character of the family including the primary state labeled by $j$.
Let us consider the partition functions with twisted boundary conditions.
First we consider the partition function $Z_{(h,\id )} = \text{Tr}  (h \ e^{-2\pi \text{Im}\tau H} e^{i 2\pi \text{Re}\tau P})$.
This partition function is given by 
\be
Z_{(h,\id )} = \sum_{j \in \frac{1}{2}\mathbb{Z}} ^{\frac{k}{2}} (-1)^{2j} |\chi_j|^2.
\ee
The modular $S$ matrix and $T$ matrix are given by 
\ba
S_{j j'} &=& \s{\f{2}{k+2}} \sin \frac{\pi (2j+1)(2j'+1)}{k+2} , \notag \\
T_{jj'} &=& e^{2\pi i (\Delta_j -\frac{c}{24}) }\delta_{j j'}, \label{eq:modularM}
\ea
where $\Delta _j  = \frac{j(j+1)}{k+2}$ and $c=\frac{3k}{k+2}$. 

Let us assume modular invariance and see if any contradiction appears.
Using (\ref{eq:modularM}), we can compute $\mcl{S}Z_{(h,\id )} = Z_{(\id ,h)}$ 
and $\mathcal{T}^l\mcl{S}Z_{(h,\id )} = Z_{(h^l,h)}$ directly.
The result is given by 
\ba
Z_{(\id ,h)} &=& \sum_{j \in \frac{1}{2}\mathbb{Z}} ^{\frac{k}{2}} \chi_{j} \bar{\chi}_{\f{k}{2}-j}, \notag\\
Z_{(h^l,h)} &=& \sum_{j \in \frac{1}{2}\mathbb{Z}} ^{\frac{k}{2}} (-i)^{kl} (-1)^{2j l} \chi_{j} \bar{\chi}_{\f{k}{2}-j}. \label{eq:twistedsu2}
\ea
Especially, we obtain $Z_{(h^2,h)} = (-1)^k Z_{(\id ,h)}$.
Since $h^2=\id $, we find a contradiction 
when $k$ is an odd number.
In other words, the theory has a mixed global 't Hooft anomaly when $k$ is an odd number.

We can compute orbifold partition function (\ref{eq:orbifold}) using (\ref{eq:twistedsu2}) when $k$ is an even number.
The results are given by\cite{FuOs}\cite{GeWi}
\be
Z_{orb} = \sum_ {\begin{subarray}{c} j=0 \\ j  \in  \mathbb{Z} \end{subarray}} ^{\lfloor \frac{k}{2} \rfloor} |\chi_j|^2  +  \sum_ {\begin{subarray}{c} j=0 \\ j  \in  \frac{1}{2}  \mathbb{Z} \end{subarray}} ^{ \frac{k}{2} } \frac{1+(-i)^k(-1)^{2j}}{2} \chi_j \bar{\chi}_{\frac{k}{2}-j}.
\ee
When $k = 4l \ (l \in \mathbb{Z})$, this gives so called $D_{2l}$ type modular invariants:
\be
Z_{orb} = \sum_ {\begin{subarray}{c} j=0 \\ j  \in  \mathbb{Z} \end{subarray}} ^{l-1} |\chi_j + \chi_{2l-j}|^2  + 2 |\chi_l|^2.
\ee
On the other hand, when $k = 4l+2$, $Z_{orb}$ becomes so called $D_{2l+1}$ type modular invariants:
\be
Z_{orb} = \sum_ {\begin{subarray}{c} j=0 \\ j  \in  \mathbb{Z} \end{subarray}} ^{2l+1} |\chi_j  |^2 + |\chi_{l+\frac{1}{2}}|^2 + \sum_ {\begin{subarray}{c} j=\f{1}{2} \\ j  \in    \mathbb{Z} + \f{1}{2} \end{subarray}} ^{ l-\f{1}{2}} ( \chi_j \bar{\chi}_{2l+1-j} + \bar{\chi}_j \chi_{2l+1-j} ).
\ee



Finally we give a comment on a difference between $k=4l$ and $k=4l+2$.
Actually, $D_{2l+1}$ type modular invariants contain half odd spins while $D_{2l}$ type modular invariants do not.
This mod $4$ structure comes from the phase factor $(-i)^k$ in $Z_{(h,h)} $. See eq.~(\ref{eq:twistedsu2}) .
This is related to the consistency of the $\SO(3)$ Chern-Simons (CS) theory\cite{DiWi}.
A level of Chern Simons theory is quantized depending on the least instanton number in $4$ dimensions\cite{DiWi}.
In $\SO(3)$ cases, the instanton number is a multiple of $\frac{1}{4}$.
Therefore, the level $k$ of an $\SO(3)$ CS theory is quantized to a multiple of $4$ in $\SU (2)$ language.
On the other hand, if we consider spin Chern-Simons theories, the quantization condition is changed and levels $k = 4l+2$ are allowed\cite{DiWi}.

\subsection{General WZW model for simple Lie group $G$} 
In this subsection we consider the WZW model for a simple Lie group $G$.
The consistency condition for the orbifold by the center of $G$ has been studied \cite{AhWa}\cite{FGK}\cite{GaSc}\cite{Gaberdiel}.
In this subsection we summarize these results.
We follow the notations and the conventions of \cite{Yellow} unless otherwise stated.

\begin{figure}
  \centering
  \begin{tabular}{ccc}
    \includegraphics{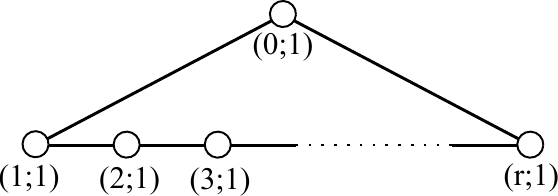} & \includegraphics{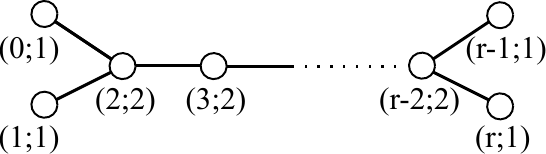}\\
    $A_r,\quad r\ge 2$ & $D_r,\quad r\ge 4$ \\
    \ &\\
    \includegraphics{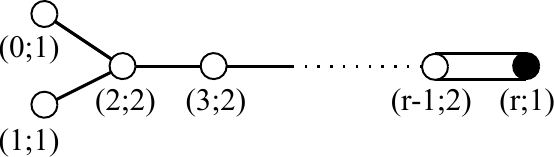} &
    \includegraphics{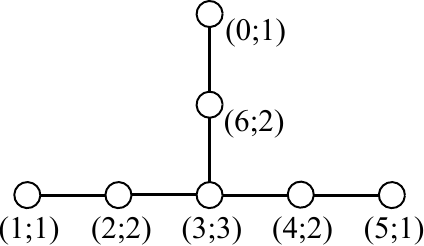}\\
    $B_r,\quad r\ge 3$ & $E_6$ \\
    \ &\\
    \includegraphics{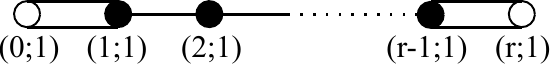} & \includegraphics{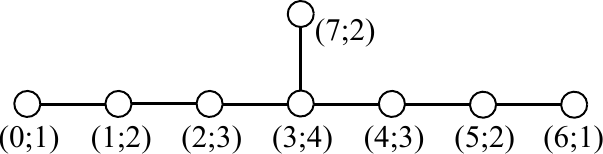}\\
    $C_r,\quad r\ge 2$ & $E_7$ 
  \end{tabular}
  \caption{The extended Dynkin diagrams and the comarks of the relevant Lie algebras.  The symbol $(i;a_{i}^{\vee})$ at each node denotes the label of the corresponding simple root and the comark. Each black node $\bullet$ corresponds to a short root, while each white node $\circ$ corresponds to a long root.}
  \label{fig:Dynkin}
\end{figure}

Let us see these conditions from the perspective of mixed global anomalies.
We consider the case that the center $\Gamma$ of $G$ is a cyclic group $\mathbb{Z}_{N}$.
In WZW models for general Lie groups $G$, the Lagrangian is given in the similar way to  (\ref{eq:LagWZWsu2}).
The primaries are labeled by a set of non-negative integers, 
called affine Dynkin labels, 
$\hat{\lambda} = [\lambda_0;\lambda_1,\cdots ,\lambda_{r}]$ 
where $r$ is the rank of $G$ with the condition.
The level $k$ is related to 
$\hat{\lambda}$ by
\be
k = \lambda_0 + \sum_{i=1}^{r}a_i^{\vee}\lambda_i,\label{level-Dynkinlabel}
\ee
where $a_i^{\vee},\ i=1,\cdots,r$ are the comarks of the corresponding Lie algebra. 
Our convention for the labels of the simple roots is the same as \cite{Yellow}. 
The labels of the simple roots and the comarks of the relevant Lie algebras are summarized in Figure \ref{fig:Dynkin}.
The non-negativity of $\lambda_0$ and the relation \eqref{level-Dynkinlabel} give a constraint on the allowed representations:
\be
0 \le \sum_{i=1}^{r}a_i^{\vee}\lambda_i \le k.
\ee
$P_+^{k}$ denotes the set of affine Dynkin labels which satisfy the above condition.

We consider the center group $\mathbb{Z}_{N} = \{\id, h, \cdots, h^{N-1}\}$ generated by $h$.
The action of $h$ is given by the following equation \cite{Yellow}:
\be
h \ket{\hat{\lambda}, \hat {\lambda}} = e^{-2 \pi i (A \hat{\omega}_0, \lambda)}\ket{\hat{\lambda}, \hat {\lambda}} . \label{eq:center}
\ee
We also use $h$ to represent the matrix whose elements are given by $h_{\hat{\mu},\hat{\lambda}} =e^{-2 \pi i (A \hat{\omega}_0, \lambda)} \delta _{\hat{\mu},\hat{\lambda}}$.

Another important operation is the action of outer automorphisms of the affine Lie algebra $\hat{\mathfrak{g}}$.
The action of the generator $A$ of outer automorphisms \footnote{Here we assume that outer automorphisms are cyclic groups.} on a character is induced from the action on the weight lattice: 
\be
A \chi_{\hat{\lambda}} = \chi_{A\hat{\lambda}}.
\label{Achi}
\ee
In the matrix notation, we can represent $A_{\hat{\mu},\hat{\lambda}} = \delta_{\hat{\mu},A\hat{\lambda}}$ where the action of matrices is defined by $A\chi_{\hat{\lambda}} = \sum_{\hat{\mu}} \chi_{\hat{\mu}}A_{\hat{\mu},\hat{\lambda}}$.
The center of $G$ is isomorphic to outer automorphisms of the corresponding affine Lie algebra $\hat{\mathfrak{g}}$.
Actually, the isomorphism is given by the modular S matrix as 
\be
S A S  ^{\dagger}= h.
\label{SAS}
\ee
In the matrix form, we can represent this as
$(S A S  ^{\dagger})_{\hat{\mu},\hat{\lambda}} = \sum_{\hat{\sigma},\hat{\rho}} S_{\hat{\lambda},\hat{\sigma}} A_{\hat{\sigma},\hat{\rho}}(S^{\dagger})_{\hat{\rho},\hat{\mu} } $ \footnote{This notation is different from \cite{Yellow}.}.
Here we introduce modular S matrix $S_{\hat{\mu},\hat{\lambda}}$ by $\mathcal{S}\chi_{\hat{\lambda}} = \sum_{\hat{\mu}} \chi_{\hat{\mu} } S_{\hat{\mu},\hat{\lambda}}$.

We employ the same strategy as $\SU(2)$ case;
we assume the modular invariance and see if any contradiction appears.
Then the twisted sector partition function (before the projection onto gauge invariant states) is given by the outer automorphism of $\hat{\mathfrak{g}}$:
\be
Z_{(\id ,h)} = \mcl{S}Z_{(h,\id )} =  \sum_{\hat{\mu}\in P_+^k} \bar{\chi}_{A\hat{\mu}} \chi_{\hat{\mu}},
\ee
where we use Eqs.~\eqref{Achi} and \eqref{SAS}.
Another formula we need is the modular $T$ transformation of $A$ where $T_{\hat{\mu},\hat{\lambda}} = e^{2\pi i (\Delta_{\hat{\mu}} - \frac{c}{24})} \delta_{\hat{\mu},\hat{\lambda}}$.
This is given by \cite{Yellow}
\be
(T^\dagger AT)_{\hat{\mu},\hat{\lambda}} = \delta_{\hat{\mu},A\hat{\lambda}} e^{-\pi i k |A\hat{\omega}_0|^2 - 2\pi i (A \hat{\omega}_0,\lambda)}.
\ee
Using these relations, the partition function $Z_{(h^l,h)}$ is given by
\be
Z_{(h^l,h)} = \mcl{T}^l\mcl{S}Z_{(h,\id )} = \sum_{\hat{\mu}\in P_+^k} e^{-\pi i kl |A\hat{\omega}_0|^2 - 2\pi il (A \hat{\omega}_0,\lambda)} \bar{\chi}_{A\hat{\mu}} \chi_{\hat{\mu}}. \label{eq:stlpartfunc}
\ee
The phase $e^{- 2\pi i (A \hat{\omega}_0,\lambda)}$ is exactly the same as that in (\ref{eq:center}).
Therefore, this phase means the action of the center $\Gamma$ and satisfies $e^{- 2\pi i N (A \hat{\omega}_0,\lambda)} = 1$ where $N$ is the order of $\Gamma$.
Then, 
by substituting $l=N$ in (\ref{eq:stlpartfunc}) we obtain
\be
Z_{(h^N,h)} = e^{-\pi i k N|A\hat{\omega}_0|^2} Z_{(\id ,h)}.
\ee
Thus if the phase $e^{-\pi i kN |A\hat{\omega}_0|^2}\ne 1$, we have contradiction and mixed global anomalies arise.

We list the values of  $e^{-\pi i N |A\hat{\omega}_0|^2}$ in table \ref{table:centerliegroups} 
for arbitrary compact, simple, connected and simply connected Lie groups 
whose center groups are non-trivial cyclic groups.
According to this list, center symmetries of $\SU (2r)$, $\Sp (2r+1)$, $\Spin (4r+2)$ and $E_7$ can be anomalous.
In these cases, only for even $k$ the center symmetries are not anomalous.
By the 't Hooft anomaly matching condition, a theory with even $k$ and a theory with odd $k$ are not connected by RG flows while preserving the center symmetry.

\begin{table}[hhh]
\centering
\begin{tabular}{|c|c|c|c|c|c|}\hline
Cartan matrix 
&  Group $G$  & center $\Gamma$ & $|A\hat{\omega}_0|^2 $ & $e^{-\pi iN |A\hat{\omega}_0|^2 }$ & Anomaly Free   \\\hline
$A_{n-1}$ & $\SU (n)$ & $\mathbb{Z}_{n}$ &  $|\omega_1|^2 = \frac{n-1}{n}$ & $(-1)^{n-1}$ & $n\in 2\mathbb{Z}+1$ or $k \in 2 \mathbb{Z}$ \\ 
$B_{n}$ & $\Spin (2n+1)$ & $\mathbb{Z}_2$ & $|\omega_1|^2 = 1$ & $1$  & $k \in \mathbb{Z}$ \\ 
$C_{n}$ & $\Sp (n)$ & $\mathbb{Z}_2$ & $|\omega_n|^2 = \frac{n}{2}$ & $(-1)^n$ & $n\in 2\mathbb{Z}$ or $k \in 2\mathbb{Z}$ \\ 
$D_{2l+1}$ & $\Spin (4l+2)$ & $\mathbb{Z}_4$ & $|\omega_1|^2 = \frac{2l+1}{2}$ & $-1$ & $k \in 2\mathbb{Z}$ \\
$E_6$ & $E_6$ & $\mathbb{Z}_3 $ & $|\omega_5|^2 = \frac{4}{3}$ & $1$ & $k \in \mathbb{Z}$ \\ 
$E_7$ & $E_7$ & $\mathbb{Z}_2 $ & $|\omega_6|^2 = \frac{3}{2}$ & $-1$ & $k \in 2\mathbb{Z}$ \\ 
\hline
\end{tabular}
\caption{Lie groups $G$ denote the simply connected counterparts of given type Lie algebras. 
$N$ is the order of their center symmetry, which are cyclic groups.
There are possibilities of global anomalies only when the phase $e^{-\pi i N |A\hat{\omega}_0|^2}$ is not equal to one.\label{table:centerliegroups}}
\end{table}

When there are no mixed gauge gravitational anomalies, we can construct the orbifold partition function by
\be
Z_{orb} =   \sum_{\hat{\mu},\hat{\lambda}\in P_+^k} \bar{\chi}_{\hat{\mu}} \mcl{M}_{\hat{\mu},\hat{\lambda}} \chi_{\hat{\lambda}},
\ee
where the matrix element $\mcl{M}_{\hat{\mu}\hat{\lambda}}$ is given by 
\be
\mcl{M}_{\hat{\mu}\hat{\lambda}} = \frac{1}{N}\sum_{p,q = 0}^{N-1}\delta_{\hat{\mu},A^p\hat{\lambda}} e^{-2\pi  i q (A\hat{\omega}_0,\lambda)}e^{-\pi i p q k |A\hat{\omega}_0|^2}.
\ee
This partition function is, of course, not invariant under the modular transformation or gauge transformation when there are mixed gauge gravitational anomalies.

\subsection{Anomalies of subgroups}\label{anomalysub}
We can also consider anomalies for subgroups of the center.
Let us consider the subgroup $\mathbb{Z}_{M} \subset \mathbb{Z}_{N}$ where $M$ satisfies $Ms = N$ for an integer $s$.
Then the generator of $\mathbb{Z}_M$ is given by $h' = h^s$ i.e. $\mathbb{Z}_M = \{1,h^s, \cdots, h^{s(M-1)}\}$.
Under the assumption of modular invariance, the twisted sector partition function $Z_{(\id ,h')} = \mcl{S}Z_{(h',\id )}$ is 
\be
Z_{(\id ,h')} = \sum_{\hat{\mu}\in P_+^k} \bar{\chi}_{A^s\hat{\mu}} \chi_{\hat{\mu}},
\ee 
and the sector $(h'^l,h')$ partition function $Z_{(h'^l,h')} = \mcl{T}^l\mcl{S}Z_{(h',\id )}$ is 
\be
Z_{(h'^l,h')} = \sum_{\hat{\mu}\in P_+^k} e^{-\pi i k l |A^s\hat{\omega}_0|^2}  e^{-2\pi i ls^2(A\hat{\omega}_0,\mu)}\bar{\chi}_{A^s\hat{\mu}}\chi_{\hat{\mu}}.
\ee
Similarly, the matrix element of orbifold $\mcl{M}_{\hat{\mu}\hat{\lambda}}$ is given by 
\be
\mcl{M}_{\hat{\mu}\hat{\lambda}} = \frac{1}{M}\sum_{p,q = 0}^{M-1}\delta_{\hat{\mu},A^{ps}\hat{\lambda}} e^{-2\pi  i qs^2 (A\hat{\omega}_0,\lambda)}e^{-\pi i p q k |A^s\hat{\omega}_0|^2}.
\ee
Anomalies are detected by the following phase factor:
\be
Z_{(h'^M,h')} = e^{- \pi i M k|A^s\hat{\omega}_0|^2}Z_{(\id ,h')}.
\ee
As an example, we consider the $\SU (6)$ $k=1$ WZW model and the subgroup $\mathbb{Z}_2$ of the center $\mathbb{Z}_6$.
In this case, $|A^3\hat{\omega}_0|^2 = |\omega_3|^2 = \frac{3}{2}$ and $e^{- \pi i 2 |A^s\hat{\omega}_0|^2} = -1$.
Therefore, this subgroup $\mathbb{Z}_2$ is still anomalous.

Another example is the $\SU (8)$ $k=1$ WZW model, where the center $\mathbb{Z}_8$ is anomalous.
Let us consider the subgroup $\mathbb{Z}_2$ of the center $\mathbb{Z}_8$.
In this case, $|A^4\hat{\omega}_0|^2 = |\omega_4|^2 = 2$ and $e^{- \pi i 2 |A^s\hat{\omega}_0|^2} = 1$.
Therefore, $\mathbb{Z}_2 = \{1,h^4\}$ is a non-anomalous subgroup of anomalous $\mathbb{Z}_8$.
Actually, the orbifold theory gives the $E_7$ $k=1$ WZW model:
\be
Z_{orb} = |\chi_{\bm{1}}+\chi_{\bm{70}}|^2 + |\chi_{\bm{28}} + \chi_{\bar{\bm{28}}}|^2,
\ee
where the decomposition of a fundamental representation $\bm{56}$ and the adjoint $\bm{133}$ of $\mathfrak{e}_7$ under $\mathfrak{su}_8$ is given by 
\ba
\bm{133} &\to& \bm{63} + \bm{70}, \\
\bm{56} &\to& \bm{28} + \bar{\bm{28}}.
\ea 

\subsection{Cancellation of mixed global anomalies in $G\times G'$ type WZW models }
Global gauge gravitational anomalies can be cancelled by taking the tensor product of two theories with 't Hooft anomalies.
The simplest example is $\SU (2)_k\times \SU (2)_1$ theory when $k$ is odd.
The action of $\mathbb{Z}_2$ is given by the diagonal way.
Twisted sector partition functions are given by
\be
\mcl{T}^l\mcl{S}Z_{(h,\id )} = \Big( \sum_ {\begin{subarray}{c} j=0 \\ j  \in \frac{1}{2} \mathbb{Z} \end{subarray}} ^{\frac{k}{2}} (-1)^{2jl}(-i)^{kl} \chi_j\bar{\chi}_{\frac{k}{2}-j}\Big)\Big( \sum_ {\begin{subarray}{c} j'=0 \\ j'  \in \frac{1}{2} \mathbb{Z} \end{subarray}} ^{\frac{1}{2}}(-1)^{2j'l}(-i)^{l}\chi_{j'}'\bar{\chi}_{\frac{1}{2}-j'}'\Big),
\ee
where $\chi$ and $\chi'$ denote the characters of first $\SU (2)_k$ and second $\SU (2)_1$ respectively. Therefore, we obtain $Z_{(h^2,h)} = \mcl{T}^2\mcl{S}Z_{(h,\id )} = (-1)^{k+1}Z_{(\id ,h)}$ and 't Hooft anomalies are cancelled when $k$ is odd.
In these cases we can actually take the orbifold of them.
After some calculations, we obtain for $k = 4l+1 (l \in \mathbb{Z})$
\be
Z_{orb} = \sum_{\begin{subarray}{c} j=0 \\ j \in \mathbb{Z}\end{subarray}} ^{\frac{k-1}{2}} 
  |\chi_j \bar{\chi}'_0 + \chi_{\frac{k}{2}-j} \bar{\chi}_{\frac{1}{2}}'|^2,
\ee
and for $k = 4l - 1(l \in \mathbb{Z})$
\be
Z_{orb} = \sum_{\begin{subarray}{c} j=0 \\ j \in \mathbb{Z}\end{subarray}} ^{\frac{k-1}{2}} 
  |\chi_j \chi'_0 + \chi_{\frac{k}{2}-j} \chi_{\frac{1}{2}}'|^2.
\ee
As a special case, the orbifold $(\SU (2)_3\times \SU (2)_1)/\mathbb{Z}_2$ gives an orbifold construction of the $k=1$ $G_2$ WZW model:
\footnote{The $G_2$ $k=1$ WZW model includes the basic representation $\chi_{\id }$
 and fundamental representation $\chi_{\bm{7}}$: $Z_{diag}^{G_2} = |\chi_{\id }|^2 + |\chi_{\bm{7}}|^2$. 
The decomposition of a fundamental representation $\bm{7}$ and the adjoint representation $\bm{14$} of $\mathfrak{g}_2$ under $\mathfrak{su}_2 \times \mathfrak{su}_2$ is given by $\bm{7} \to (\bm{3},\id )+(\bm{2},\bm{2})$ and $\bm{14} \to (\bm{3},\id )+(\id ,\bm{3})+(\bm{4},\bm{2})$.}
\be
Z_{orb} = |\chi_0\chi'_0 + \chi_{\frac{3}{2}}\chi'_{\frac{1}{2}}|^2 + |\chi_{1}\chi'_0 + \chi_{\frac{1}{2}}\chi'_{\frac{1}{2}}|^2.
\ee

We can also consider the anomaly cancellation between WZW models of different groups.
When we consider the anomaly of diagonal $\mathbb{Z}_N$ in $G\times G'$, we can see whether there are anomalies or not by studying the anomalous phases
\be
\mcl{T}^NZ_{(\id ,h)} ^{G\times G'}= e^{-\pi i k N |A\hat{\omega}_0|^2} e^{-\pi i k' N |A\hat{\omega}_0'|^2}Z_{(\id ,h)}^{G\times G'}.
\ee
For example, we can consider the diagonal $\mathbb{Z}_2$ in $\Sp(3)_1\times \Sp(1)_1$.
In this case, the anomalous phase factor becomes $e^{-\pi i  2 |A\hat{\omega}_0|^2}e^{-\pi i 2 |A\hat{\omega}_0'|^2}=(-1)\times(-1) = 1$ and there are no global mixed anomalies.
Actually, the partition function of the orbifold CFT $(\Sp(3)_1\times \Sp(1)_1)/\mathbb{Z}_2$ 
\footnote{Representations of the $k= 1$ $\Sp(3)$ WZW model include a fundamental representation $\bm{6}$, the irreducible part of rank 2 anti symmetric tensor $\bm{14}$ and the irreducible part of rank 3 anti symmetric tensor $\bm{14}'$.
The diagonal partition function of the $k= 1$ $\Sp(3)$ WZW model is given by $Z_{diag}^{\Sp(3)_1} = |\chi_{\bm{1}}|^2 + |\chi_{\bm{6}}|^2 + |\chi_{\bm{14}}|^2 + |\chi_{\bm{14'}}|^2$. }
 is
\be
Z_{orb} = |\chi_{\bm{1}}\chi_{\bm{1}}' + \chi_{\bm{14'}}\chi_{\bm{2}}'|^2 + |\chi_{\bm{14}}\chi_{\bm{1}}' + \chi_{\bm{6}}\chi'_{\bm{2}}|^2,
\ee
which gives an orbifold realization of the $k=1$ $F_4$ WZW model
\footnote{
The $F_4$ $k=1$ WZW model includes the basic representation $\chi_{\id }$
 and fundamental representation $\chi_{\bm{26}}$: $Z_{diag}^{F_4} = |\chi_{\id }|^2 + |\chi_{\bm{26}}|^2$. 
The decomposition of a fundamental representation $\bm{26}$ and the adjoint representation $\bm{52}$ of $\mathfrak{f}_4$ under $\mathfrak{usp}_3 \times \mathfrak{usp}_1$ is given by $\bm{26} \to (\bm{14},\bm{1})+(\bm{6},\bm{2})$ and $\bm{52} \to (\bm{21},\bm{1})+(\bm{1},\bm{3}) + (\bm{14'},\bm{2})$.
}.

As another example, we can consider $\SU (2m)_1\times \SU (2)_1$ WZW models and action of $\mathbb{Z}_{2m}$ given by 
\be
\mathbb{Z}_{2m} = \{(\bm{1},\bm{1}),(h_1,h_2),(h_1^2,\bm{1}),\cdots, (h_1^{2m-1},h_2)\} \subset \mathbb{Z}_{2m}\times \mathbb{Z}_2,
\ee 
where $h_1$ and $h_2$ are the generators of the centers of $\SU (2m)$ and $\SU (2)$ respectively.
This $\mathbb{Z}_{2m}$ is generated by $h = (h_1,h_2)$.
In this case, the twisted sector partition functions $Z_{(h^l,h)}$ are
\be
Z_{(h^l,h)} = \Big(\sum_{\hat{\mu}\in P_+^1} e^{-\pi i  l |A\hat{\omega}_0|^2} h(\hat{\mu})^l \bar{\chi}_{A\hat{\mu}}\chi_{\hat{\mu}} \Big)\Big( \sum_ {\begin{subarray}{c} j=0 \\ j \in \frac{1}{2} \mathbb{Z} \end{subarray}} ^{\frac{1}{2}}(-i)^{l}(-1)^{2jl} \bar{\chi}_{\frac{1}{2}-j} \chi_{j} \Big ),
\ee
where $h(\hat{\mu}) = e^{-2\pi i(A\hat{\omega}_0,\mu)}$ is the action of the generator of the center of $\SU(2m)$ on representation $\hat{\mu}$
. 
The anomalous phase is given by $Z_{(h^{2m},h)} = e^{ \pi i (2m-1)}(-1)^{m}Z_{(\bm{1},h)}$.
Therefore, only for odd $m$ the mixed anomalies are cancelled in this manner.

\section{Boundary states and 't Hooft anomalies}\label{sec:BCFT}
In this section we consider the symmetry properties of boundary states. We find the condition of the existence of a symmetry invariant boundary state and compare it to the consistency condition of the orbifold theory.

Let us first quickly review the boundary states of WZW models.  The basis of the boundary states are so-called ``Ishibashi states'' $\ishibashi{\hat{\lambda}},\ \hat{\lambda}\in P_+^k$ \cite{Ishibashi,Onogi}, which are certain formal linear combinations of the primary states $\ket{\hat{\lambda},\hat{\lambda}}$ and their descendants.  The set of physically realized boundary states are ``Cardy states'' \cite{Cardy} denoted by $\cardy{\hat{\mu}},\ \hat{\mu}\in P_{+}^k$ and expressed by using the modular S matrix as
\begin{align}
  \cardy{\hat{\mu}} = \sum_{\hat{\lambda}\in P_+^k}
  \frac{S_{\hat{\mu},\hat{\lambda}}}{\sqrt{S_{\hat{0},\hat{\lambda}}}} \ishibashi{\hat{\lambda}},
\end{align}
where $\hat{0}=[k;0,\cdots,0]$.

Next we consider the action of the center symmetry on the Cardy states.
This action is determined through Ishibashi states.
The action of the center symmetry on Ishibashi states is read off from eq.~\eqref{eq:center}:
\be
h \ishibashi{\hat{\lambda}} 
= \sum_{\hat{\nu}\in P_+^k} h_{\hat{\nu},\hat{\lambda}}\ishibashi{\hat{\nu}}.
\ee
Then, we can determine the action on Cardy states\cite{MMS}:
\ba
h\cardy{\hat{\mu}} &=& \sum_{\hat{\lambda}\in P_+^k}
 \frac{S_{\hat{\mu},\hat{\lambda}}}{\sqrt{S_{\hat{0},\hat{\lambda}}}} h \ishibashi{\hat{\lambda}}
\notag \\ 
&=& \cardy{A\hat{\mu}},
\ea
where we used eq.~(\ref{SAS}).
Therefore, we find that the action of the center symmetry is given by the outer automorphism.

We also consider the action of the charge conjugation operator $C$ on Cardy states.
In the matrix form, this is given by $C _{\hat{\lambda},\hat{\mu}} = \delta _{C\hat{\lambda},\hat{\mu}} =\delta _{\hat{\lambda},C\hat{\mu}} $.
The charge conjugation operator is also obtained by the square of modular S matrix: $S^2 = C$.
Thus the relation $CS = S^3 = SC$ holds.
The action of the charge conjugation is also determined through the Ishibashi states:
\be
C\ishibashi{\hat{\lambda}} = \ishibashi{C\hat{\lambda}} 
= \sum_{\hat{\rho}\in P_+} C_{\hat{\rho},\hat{\lambda}}\ishibashi{\hat{\rho}}.
\ee
From this action on the Ishibashi states, we can derive the action on Cardy states:
\ba
C\cardy{\hat{\mu}}&=& \sum_{\hat{\lambda}\in P_+^k} \frac{S_{\hat{\mu},\hat{\lambda}}}{\sqrt{S_{\hat{0},\hat{\lambda}}}} C \ishibashi{\hat{\lambda}}
\notag \\
&=& \cardy{C\hat{\mu}}.
\ea
Therefore, the charge conjugation on a Cardy state is given by the charge conjugation on the Dynkin label.

\subsection{$\SU(2)_k$ cases}
We consider $\SU(2)_k$ cases. In the diagonal theories, there are $k+1$ boundary states $\cardy{j}$ that are labelled by spin $j$.
Under the center symmetry $\mathbb{Z}_2$, $\cardy{j}$ is mapped to $h\cardy{j} = \cardy{\frac{k}{2}-j}$.
Now we consider the existence of an invariant boundary state under the center symmetry.
Then, an invariant boundary state satisfies $j = \frac{k}{2} - j$, which reads $j = \frac{k}{4}$.
Because $j \in \frac{1}{2}\mathbb{Z}$, $k$ is an even integer if there exists an invariant boundary state.
This even/odd classification is the same condition that is obtained from the modular invariance of orbifold theories.
On the other hand, when $k$ is an odd integer, the system is supposed to be a non-trivial SPT phase in critical systems, and $\cardy{j}$ and $\cardy{\frac{k}{2}-j}$ always form a pair.

\subsection{$\SU (3)_1$ cases}
First, we consider the $\SU (3)_1$ WZW model, whose center $\mathbb{Z}_3$ is non-anomalous.
The modular $S$ matrix for $\SU (3)_1$ is given by 
\be
S = \frac{1}{\sqrt{3}}
\begin{pmatrix}
1 & 1 & 1 \\
1 & \kappa & \kappa ^2 \\
1 & \kappa ^2 &\kappa
\end{pmatrix}
,\quad \kappa = e^{\frac{2\pi i}{3}}.
\ee
The diagonal partition function is given by 
\be
Z_{diag} = |\chi_{\bm{1}}|^2 + |\chi_{\bm{3}} |^2 + |\chi_{\bar{\bm{3}}}|^2.
\ee
Here we used the representations of $\SU (3)$ as labels of primary fields.
Because $\mathbb{Z}_3$ is non-anomalous, we can take the orbifold of this diagonal models by the center symmetry $\mathbb{Z}_3$, and we obtain the following partition function:
\be 
Z_{orb} = |\chi_{\bm{1}}|^2 + \chi_{\bm{3}}\bar{\chi}_{\bar{\bm{3}}} + \chi_{\bar{\bm{3}}}\bar{\chi}_{\bm{3}}.
\ee
The orbifold action exchanges the $\bm{3}$ and $\bar{\bm{3}}$.

By the action of the center symmetry, boundary states are interchanged cyclically: 
\ba
h\cardy{\bm{1}} &=& \cardy{\bm{3}},  \notag \\
h\cardy{\bm{3}} &=& \cardy{\bar{\bm{3}}},  \notag \\
h\cardy{\bar{\bm{3}}} &=& \cardy{\bm{1}}.
\ea
Therefore, there are no boundary states invariant under the center symmetry.
This seems to suggest that the boundary state approach does not work in this case.

Alternatively, let us consider the combination $Ch$. 
This combination satisfies $(Ch)^2 = 1$ and generates $\mathbb{Z}_2$ symmetry.
This $\mathbb{Z}_2$ is included in $\mathbb{Z}_3 \rtimes \mathbb{Z}_2$ which is the semidirect product of the center $\mathbb{Z}_3$ and the charge conjugation $\mathbb{Z}_2$.
The action of $Ch$ to the Cardy states is given by 
\ba
Ch\cardy{\bm{1}} &=& \cardy{\bar{\bm{3}}},  \notag \\
Ch\cardy{\bm{3}} &=& \cardy{\bm{3}},  \notag \\
Ch\cardy{\bar{\bm{3}}} &=& \cardy{\bm{1}}.
\ea
Therefore, $\cardy{\bm{3}}$ is invariant under the $\mathbb{Z}_2$ generated by $Ch$.
Though the reason to introduce $C$ is mysterious, we find that the condition for the existence of a $Ch$ invariant boundary state reproduces the same condition for the modular invariance of the orbifold theory by the center $\mathbb{Z}_3$. 

This result motivates us to consider a  $Ch$ invariant boundary state in WZW models for other groups.

\subsection{General WZW models for simple groups}
\subsubsection{$A_{n-1}$ cases}
Let us consider the $\SU (n)_k$ WZW models. 
As we showed, the generator of the center symmetry $h$ acts as the outer automorphism on boundary states $h\cardy{\hat{\mu}} = \cardy{A\hat{\mu}}$.
When we use the affine Dynkin labels $ [\lambda_0;\lambda_1 \cdots \lambda_{n-1}]$ to represent $\hat{\lambda}$, the action of the outer automorphism is expressed as 
\be
A : [\lambda_0;\lambda_1, \cdots,\lambda_{n-2} ,\lambda_{n-1}] \to [\lambda_{n-1};\lambda_0,\lambda_1, \cdots , \lambda_{n-2}].
\ee
In other words, the outer automorphism interchanges the Dynkin labels cyclically. 
Then, the existence of invariant boundary states 
is equivalent to the existence of invariant affine Dynkin labels 
by this transformation.
When $[\lambda_0;\lambda_1, \cdots ,\lambda_{n-1}] = [\lambda_{n-1};\lambda_0, \cdots ,\lambda_{n-2}]$ is satisfied, all labels $\lambda_i$ must have the same value.
The relation \eqref{level-Dynkinlabel} for $\SU(n)$ reads
\be
k = \lambda_0 + \lambda_1 + \cdots + \lambda_{n-1},
\ee 
since $a_i^{\vee}=1,\ i=1,2,\dots,n-1$ for $\SU(n)$.
When there exists an invariant boundary state under the center symmetry, 
the above relation implies $k = n\lambda_0$ 
and therefore $k$ is a multiple of $n$.
This condition is not the same one as the orbifold analysis, where we can take the orbifold when $k$ is even for even $n$ and $k$ is arbitrary for odd $n$.

However, when we consider the combined symmetry $Ch$, we obtain the same results as the orbifold analysis as follows.
The action of $Ch$ on boundary states is given by 
$Ch\cardy{\hat{\mu}}=\cardy{CA \hat{\mu}}$.
The action of charge conjugation $C$ is given by the flip of Dynkin labels:
\be
C : [\lambda_0;\lambda_1,\lambda_2  ,\cdots,\lambda_{n-2} ,\lambda_{n-1}] \to [\lambda_0; \lambda_{n-1},\lambda_{n-2},  \cdots,\lambda_2 ,\lambda_1].
\ee
Therefore, under the combined action of $C$ and $A$, affine Dynkin labels are flipped as follows:
\be
CA : [\lambda_0;\lambda_1,\lambda_2 ,\cdots,\lambda_{n-2} ,\lambda_{n-1}] \to [ \lambda_{n-1};\lambda_{n-2}, \cdots ,\lambda_2, \lambda _1, \lambda_0].\label{CA}
\ee
Let us consider whether there exists an invariant boundary state.

For odd $n$, $\lambda_{\frac{n-1}{2}}$ is fixed under this $CA$ action \eqref{CA}.
Thus, if $\hat{\lambda}$ is invariant under the CA action \eqref{CA}, the relation  $\lambda_{i}=\lambda_{n-1-i}$ holds and the level $k$ is given by
\ba
k &=& \lambda_0 + \cdots + \lambda_{n-1}  \notag \\
&=& 2(\lambda_0+ \cdots + \lambda_{\frac{n-3}{2}}) + \lambda_{\frac{n-1}{2}}.
\ea
Therefore, for any value of $k$ we find an invariant boundary state.
This result is the same as that obtained by the orbifold analysis, where we found that for any value of $k$ the center symmetry  $\mathbb{Z}_n$ is non-anomalous for odd $n$.

On the other hand, when $n$ is an even integer, if $\hat{\lambda}$ is invariant under $CA$ \eqref{CA}, the value of $k$ is given by
\ba
k &=& \lambda_0 + \cdots + \lambda_{n-1}  \notag \\
 &=& 2(\lambda_0+ \cdots + \lambda_{\frac{n-2}{2}}).
\ea 
This implies that $k$ must be a multiple of $2$ 
when there exists an invariant boundary state.
As we showed in the orbifold analysis, for even $k$ the center symmetry is not anomalous and we can take the orbifold.
Therefore, the existence of an invariant boundary state under the action of $Ch$ gives the same condition for the absence of the mixed global anomalies.

\begin{table}[htb]
\centering
\begin{tabular}{|c|ccc|}\hline
Cartan matrix & affine Dynkin label &   & combined action of $C$ and $A$  \\\hline
$A_{n-1}$ & $[\lambda_0;\lambda_1,\lambda_2,\cdots,\lambda_{n-2}, \lambda_{n-1}]$ & $\to$  & $[\lambda_{n-1};\lambda_{n-2},\cdots,\lambda_2,\lambda_{1},\lambda_0]$ \\ 
$B_{n}$ & $[\lambda_0;\lambda_1,\lambda_2,\cdots,\lambda_{n-1},\lambda_n]$& $\to$   & $[\lambda_1;\lambda_0,\lambda_2,\cdots,\lambda_{n-1},\lambda_n]$  \\ 
$C_{n}$ & $[\lambda_0;\lambda_1,\cdots,\lambda_{n-1},\lambda_n]$ & $\to$  & $[\lambda_n;\lambda_{n-1},\cdots,\lambda_1,\lambda_0]$ \\ 
$D_{2l+1}$ & $[\lambda_0;\lambda_1,\lambda_2,\cdots,\lambda_{2l-1},\lambda_{2l},\lambda_{2l+1}]$& $\to$ & $[\lambda_{2l};\lambda_{2l+1},\lambda_{2l-1},\cdots,\lambda_2,\lambda_0,\lambda_1]$\\  
$E_6$ & $[\lambda_0;\lambda_1,\lambda_2,\lambda_3,\lambda_4,\lambda_5,\lambda_6]$& $\to$ & $[\lambda_1;\lambda_0,\lambda_6,\lambda_3,\lambda_4,\lambda_5,\lambda_2]$ \\ 
$E_7$ & $[\lambda_0;\lambda_1,\lambda_2,\lambda_3,\lambda_4,\lambda_5,\lambda_6,\lambda_7]$ & $\to$ & $[\lambda_6;\lambda_5,\lambda_4,\lambda_3,\lambda_2,\lambda_1,\lambda_0,\lambda_7]$\\ 
 \hline
\end{tabular}\caption{This table shows the combined action of the charge conjugate and the outer automorphisms.
For type $B_n$, $C_n$, and $E_7$ the complex conjugate acts trivially. }
\end{table}

\subsubsection{$B_n$ cases}
The condition for an invariant boundary state is 
\be
[\lambda_0 ; \lambda_1 , \lambda_2 , \cdots ,\lambda_{n-1}, \lambda_n] = [\lambda_1 , \lambda_0 , \lambda_2 , \cdots , \lambda_{n-1},\lambda_n].
\ee
Therefore, the affine Dynkin labels of an invariant boundary state satisfy 
\be
\lambda_0=\lambda_1.
\ee
The comarks for $B_n$ are given by
\be
(a_0^\vee;a_1^\vee,a_2^\vee,\cdots, a_{n-1}^\vee,a_n^\vee) = (1;1,2,\cdots,2,1). 
\ee
Then, when there is an invariant boundary state, $k$ satisfies \eqref{level-Dynkinlabel}:
\be
k = \lambda_0+\sum_{i=1}^n a_i^\vee \lambda_i 
= 2(\lambda_1 + \lambda_2+\cdots + \lambda_{n-1}) + \lambda_n.
\ee
Because $\lambda_n$ can take an arbitrary non-negative integer, we can always find an invariant boundary state at $[0;0,\cdots,0 ,k]$.
This result is the same as the consistency of the orbifold by the center $\mathbb{Z}_2$: for arbitrary $k$ we can take the orbifold.
Therefore, for $B_n$ the existence of an invariant boundary state gives the same condition for the absence of the mixed global anomalies.

\subsubsection{$C_n$ cases}
The condition for an invariant boundary state is 
\be
[\lambda_0 ; \lambda_1 , \cdots ,\lambda_{n-1}, \lambda_n] = [\lambda_n ; \lambda_{n-1} , \cdots ,\lambda_1, \lambda_0].
\ee
Therefore, the affine Dynkin labels of an invariant Boundary state satisfy
\begin{align}
  \lambda_i = \lambda_{n-i},\quad i = 0 , \dots, n . \label{invlabelC}
\end{align}
The comarks for $C_n$ is given by
\be
(a_0^\vee;a_1^\vee,\cdots, a_{n-1}^\vee,a_n^\vee) = (1;1,\cdots,1,1).
\ee

When $n$ is even and eq.~\eqref{invlabelC} is satisfied, eq.~\eqref{level-Dynkinlabel} reads
\be
k = 2(\lambda_0 + \cdots +\lambda_{\frac{n-2}{2}}) + \lambda_{\frac{n}{2}}.
\ee
Because $\lambda_{\frac{n}{2}}$ can take any positive interger, we always find an invariant boundary state at the Dynkin labels $[0;0,\cdots ,0,\lambda_{\frac{n}{2}}=k,0,\cdots ,0]$.

When $n$ is odd and eq.~\eqref{invlabelC} is satisfied, eq.~\eqref{level-Dynkinlabel} reads
\be
k = 2(\lambda_0 + \cdots + \lambda_{\frac{n-1}{2}}).
\ee
Therefore, only when $k$ is even we can find an invariant boundary state.

The above condition is the same one as the consistency of the orbifold by the center $\mathbb{Z}_2$: for even $n$ we can take the orbifold for arbitrary $k$ and for odd $n$ we can do that only when $k$ is even.
Therefore, for $C_n$ the existence of an invariant boundary state gives the same condition for the absence of the mixed global anomalies.

\subsubsection{$D_{2l+1}$ case}
The condition for an invariant boundary state under $Ch$ is 
\be
[\lambda_0;\lambda_1,\lambda_2,\cdots,\lambda_{2l-1},\lambda_{2l},\lambda_{2l+1}] = [\lambda_{2l};\lambda_{2l+1},\lambda_{2l-1},\cdots,\lambda_2,\lambda_0,\lambda_1].
\ee
Therefore, the affine Dynkin labels of an invariant boundary state satisfy 
\be
\lambda_0 = \lambda_{2l}, \ \  \lambda_1 = \lambda_{2l+1}, \ \ \lambda_i = \lambda_{2l+1-i}\ \ (\text{for} \ \ 2\le i \le l).
\label{invlabelD}
\ee
The comarks are given by
\be
(a_0^\vee; a_1^\vee, a_2 ^\vee, \cdots , a_{2l-1}^\vee,a_{2l}^\vee, a_{2l+1}^\vee) = (1,1,2,\cdots ,2 ,1,1).
\ee 
Then, when there is an invariant boundary state, $k$ satisfies \eqref{level-Dynkinlabel}:
\be
k = 2(\lambda_0+\lambda_1) + 4 (\lambda_2 + \cdots + \lambda_l).
\ee
Therefore, if there exists an invariant  boundary state, $k$ is an even integer.
This condition is the same one as the consistency of the orbifold by the center $\mathbb{Z}_4$: for arbitrary $n$ we can take the orbifold only when $k$ is even.
Therefore, for $D_{2l+1}$ the existence of an invariant boundary state gives the same condition for the absence of the mixed global anomalies.

\subsubsection{$E_6$ case}
The condition for an invariant boundary state under $Ch$ is
\be
[\lambda_0;\lambda_1,\lambda_2,\lambda_3,\lambda_4,\lambda_5,\lambda_6] = [\lambda_1;\lambda_0,\lambda_6,\lambda_3,\lambda_4,\lambda_5,\lambda_2].
\ee
Therefore, the affine Dynkin labels of an invariant boundary state satisfy
 \be
\lambda_0 = \lambda_1, \ \ \lambda_2 = \lambda_6.
\ee
The comarks are given by 
\be
(a_0^\vee;a_1^\vee,a_2^\vee,a_3^\vee,a_4^\vee, a_5^\vee, a_6^\vee) = (1;1,2,3,2,1,2).
\ee
Then, when there is an invariant boundary state, $k$ satisfies \eqref{level-Dynkinlabel}:
\be
k = 2\lambda_1+4\lambda_2+3\lambda_3 + 2 \lambda_4 + \lambda_5.
\ee
For example, the boundary state labelled by $[0;0,0,0,0,k,0]$ is invariant under the action of $Ch$.
Therefore, an invariant boundary state exists for arbitrary $k$.
This result is the same as the consistency of the orbifold by the center $\mathbb{Z}_3$: for arbitrary $k$ we can take the orbifold.
Therefore, for $E_6$ the existence of an invariant boundary state gives the same condition for the absence of the mixed global anomalies.

\subsubsection{$E_7$ case}
The condition for an invariant boundary state is
\be
[\lambda_0;\lambda_1,\lambda_2,\lambda_3,\lambda_4,\lambda_5,\lambda_6,\lambda_7] = [\lambda_6;\lambda_5,\lambda_4,\lambda_3,\lambda_2,\lambda_1,\lambda_0,\lambda_7].
\ee
Therefore, the affine Dynkin labels of an invariant boundary state satisfy
\be
\lambda_0 = \lambda_6, \ \ \ \lambda_1 = \lambda_5, \ \ \ \lambda_2 = \lambda_4.
\ee
The comarks for $E_7$ are given by
\be
(a_0^\vee;a_1^\vee,a_2^\vee,a_3^\vee,a_4^\vee, a_5^\vee, a_6^\vee, a_7^\vee) = (1;2,3,4,3,2,1,2).
\ee
Then, when there is an invariant boundary state, $k$ satisfies \eqref{level-Dynkinlabel}:
\be
k = 2\lambda_0+4\lambda_1+6\lambda_2 +  4 \lambda_3 + 2\lambda_7.
\ee
This means that we can only find an invariant boundary state when the level $k$ is even.

This condition is the same one as the consistency of the orbifold by the center $\mathbb{Z}_2$: we can take the orbifold only when $k$ is even.
Therefore, for $E_7$ the existence of an invariant boundary state gives the same condition for the absence of the mixed global anomalies.


\subsection{Subgroup of the center}
We can also consider the invariant boundary states under subgroups of the center.
To compare with the orbifold analysis, we consider two examples:
\begin{enumerate}
\item[(i)]
the $k = 1$ $\SU (6)$ WZW model and $\mathbb{Z}_2$ subgroup  of the center $\mathbb{Z}_6$.
\item[(ii)]
the $k = 1$ $\SU (8)$ WZW model and $\mathbb{Z}_2$ subgroup  of the center $\mathbb{Z}_8$.
\end{enumerate}

\subsubsection{$k = 1$ $\SU (6)$ WZW model and $\mathbb{Z}_2$ subgroup  of the center $\mathbb{Z}_6$}
The labels of primaries are given by $\bm{1}$, $\bm{6}$, $\bm{15}$, $\bm{20}$, $\bar{\bm{15}}$ and $\bar{\bm{6}}$.
We can compute the action of combination of $C$ and $h'$ (the generator of $\mathbb{Z}_2$ ) as follows:
\ba
&&Ch'\cardy{\bm{1}} = \cardy{\bm{20}}, \ \ \  Ch'\cardy{\bm{20}} = \cardy{\bm{1}},\notag   \\
&&Ch'\cardy{\bm{6}} = \cardy{\bm{15}}, \ \ \  Ch'\cardy{\bm{15}} = \cardy{\bm{6}}, \notag \\
&&Ch'\cardy{\bar{\bm{6}}} = \cardy{\bar{\bm{15}}}, \ \ \  Ch'\cardy{\bar{\bm{15}}} = \cardy{\bar{\bm{6}}}.
\ea
Therefore, there are no invariant boundary states.
Because in this case we found the mixed 't Hooft anomalies in the orbifold analysis in subsection \ref{anomalysub}, this results is consistent with our proposal:
the existence of an invariant boundary state gives the same condition for the absence of the mixed global anomalies.

\subsubsection{$k = 1$ $\SU (8)$ WZW model and $\mathbb{Z}_2$ subgroup  of the center $\mathbb{Z}_8$}
The labels of primaries are given by $\bm{1} $, $\bm{8}$, $\bm{28}$, $\bm{56}$, $\bm{70}$ $\bar{\bm{56}}$, $\bar{\bm{28}}$ and $\bar{\bm{8}}$.
We can compute the action of combination of $C$ and $h'$ (the generator of $\mathbb{Z}_2$) as follows:
\ba
&&Ch'\cardy{\bm{1}} = \cardy{\bm{70}}, \ \ \  Ch'\cardy{\bm{70}} = \cardy{\bm{1}}, \notag  \\
&&Ch'\cardy{\bm{8}} = \cardy{\bm{56}}, \ \ \  Ch'\cardy{\bm{56}} = \cardy{\bm{8}}, \\
&&Ch'\cardy{\bm{28}} = \cardy{\bm{28}}, \\
&&Ch'\cardy{\bar{\bm{8}}} = \cardy{\bar{\bm{56}}}, \ \ \  Ch'\cardy{\bar{\bm{56}}} = \cardy{\bar{\bm{8}}}, \\
&&Ch'\cardy{\bar{\bm{28}}} = \cardy{\bar{\bm{28}}}.
\ea
Therefore, we find invariant boundary states $\cardy{\bm{28}}$ and $\cardy{\bar{\bm{28}}}$ .
This also matches with our proposal because the orbifold by this $\mathbb{Z}_2$ subgroup is modular invariant.
On the other hand, if we only consider the action of $h' \in \mathbb{Z}_2$ , there are no invariant boundary states.
Only if we include the action of $C$, the existence of symmetry invariant boundary states gives the same condition for the absence of the mixed global anomalies.


\section{Conclusion}
In this paper, we considered the WZW model for a simply connected Lie group $G$.
We considered the mixed 't Hooft anomalies between large diffeomorphisms of a torus and the center $\Gamma$ of $G$.
We first considered the anomalies by studying the modular invariance of the orbifold explicitly.
We also showed explicitly that in some cases 't Hooft anomalies cancel in tensor products of two CFTs.
Next, we considered the boundary CFT approach:
we investigated the symmetry action on boundary states and the existence of invariant boundary states.
As a symmetry action, we consider the combined action of the charge conjugation $C$ and generator of $\Gamma$.
We obtained the condition that there exists a symmetry invariant boundary state.
Surprisingly, we found that the existence of an invariant boundary state gives the same condition for the absence of the mixed global anomalies.
If we do not include $C$, those two conditions are different in some cases.
We also confirmed that this is also true for subgroups of the center.


There are several future problems.
One of them is to consider the boundary CFT approach for $G\times G'$ WZW models.
In the orbifold analysis, we find that 't Hooft anomalies can cancel in $G\times G'$ WZW models even when both $G$ and $G'$ WZW models have 't Hooft anomalies.
It is an interesting problem to reproduce the same conditions for the cancellation of 't Hooft anomalies from the BCFT approach.
Usually, anomalies or SPT phases form a group and cancellation means that they are the inverse of each other in this group\cite{McGreevy}.
Therefore, the goal of this problem is to find an appropriate group structure in boundary states. 

We can also consider the condition $h\cardy{B_a} = \cardy{B_a}$ without the charge conjugation $C$.
In $G = \SU (n)$ cases, this condition impose that $k$ is a multiple of $n$ i.e. $k = nm$ for some $m \in \mathbb{Z}$.
One may think that this condition for the level $k$ gives a quantization condition of the level in $G/{\Gamma}$ Chern-Simons theory.
This expectation is not correct, because the level of $G/{\Gamma}$ Chern-Simons theory is a multiple of $2n$ \cite{DiWi}.
As noted in section \ref{SU(2)section}, in $\SU (2)$ case $k = 2m$ corresponds to the level that admits a spin Chern-Simons theory. 
Therefore, one possibility is that the condition $h\cardy{B_a} = \cardy{B_a}$ matches with the one for levels that admit spin Chern-Simons theories in $2+1$ dimensions, which is a natural generalization of the $G = SU(2)$ case.
It is interesting future problem to confirm this. 
Another interesting possibility is that this $\mathbb{Z}_n$ classification could be related to $(2+1)$-dimensional SPT phases  with $\SU(n)/\mathbb{Z}_n$ symmetry \cite{Duivenvoorden:2012yr}\cite{Roy:2015ars}.

It is also interesting to consider a connection between quantum entanglement or tensor networks.
In the case of SPT phases,
one can also distinguish the phases through the imaginary cut accompanied to spatial entanglement
even without putting the theory on manifolds with actual boundaries.
In this case, the entanglement spectrum shows non-trivial degeneracies 
when the theory is in a non-trivial SPT phase \cite{PTBO}\cite{TaTo}.
Let us see the 't Hooft anomalies of $1+1$ dimensions as a classification of a critical phase analog of SPT phases\cite{FuOs}.
We showed that we can see 't Hooft anomalies through putting CFTs on manifolds with boundaries.
Then, the natural guess is that we can also see 't Hooft anomalies through the entanglement spectrum.
We can relate boundary CFTs to the entanglement spectrum in CFTs\cite{CaTo}\cite{ACT}.
The connection of Multi Entanglement Renormalization Ansatz (MERA), which describes the critical point ground state wave function, and 't Hooft anomalies of purely internal symmetries are discussed in \cite{BrWi}.
There are also MERA with symmetries\cite{SiVi} or with boundaries\cite{EveVi}.
It is interesting to study the relation among them.







\section*{Acknowledgements}
We would like to thank
Masaki Oshikawa, Ken Shiozaki, Tadashi Takayanagi, Shinsei Ryu, Hidenori Fukaya and Tetsuya Onogi
for helpful discussions.
TN is supported by JSPS fellowships and the Simons Foundation.
SY is supported in part by JSPS KAKENHI Grant Number JP15K05054.

\bibliography{anomalybunken} 

\providecommand{\href}[2]{#2}\begingroup\raggedright\begin{thebibliography}{10}

\bibitem{tHooft}
G.~'t~Hooft, ``{Naturalness, chiral symmetry, and spontaneous chiral symmetry
  breaking},''
\href{http://dx.doi.org/10.1007/978-1-4684-7571-5_9}{{\em NATO Sci. Ser. B}
  {\bfseries 59} (1980) 135--157}.

\bibitem{CsMu}
C.~Csaki and H.~Murayama, ``{Discrete anomaly matching},''
  \href{http://dx.doi.org/10.1016/S0550-3213(97)00839-0}{{\em Nucl. Phys.}
  {\bfseries B515} (1998) 114--162},
\href{http://arxiv.org/abs/hep-th/9710105}{{\ttfamily arXiv:hep-th/9710105
  [hep-th]}}.

\bibitem{CGLW}
X.~Chen, Z.-C. Gu, Z.-X. Liu, and X.-G. Wen, ``{Symmetry protected topological
  orders and the group cohomology of their symmetry group},''
  \href{http://dx.doi.org/10.1103/PhysRevB.87.155114}{{\em Phys. Rev.}
  {\bfseries B87} no.~15, (2013) 155114},
\href{http://arxiv.org/abs/1106.4772}{{\ttfamily arXiv:1106.4772
  [cond-mat.str-el]}}.

\bibitem{CaHa}
C.~G. Callan, Jr. and J.~A. Harvey, ``{Anomalies and Fermion Zero Modes on
  Strings and Domain Walls},''
\href{http://dx.doi.org/10.1016/0550-3213(85)90489-4}{{\em Nucl. Phys.}
  {\bfseries B250} (1985) 427--436}.

\bibitem{KaTh1}
A.~Kapustin and R.~Thorngren, ``{Anomalies of discrete symmetries in three
  dimensions and group cohomology},''
  \href{http://dx.doi.org/10.1103/PhysRevLett.112.231602}{{\em Phys. Rev.
  Lett.} {\bfseries 112} no.~23, (2014) 231602},
\href{http://arxiv.org/abs/1403.0617}{{\ttfamily arXiv:1403.0617 [hep-th]}}.

\bibitem{KaTh2}
A.~Kapustin and R.~Thorngren, ``{Anomalies of discrete symmetries in various
  dimensions and group cohomology},''
\href{http://arxiv.org/abs/1404.3230}{{\ttfamily arXiv:1404.3230 [hep-th]}}.

\bibitem{GKSW}
D.~Gaiotto, A.~Kapustin, N.~Seiberg, and B.~Willett, ``{Generalized Global
  Symmetries},'' \href{http://dx.doi.org/10.1007/JHEP02(2015)172}{{\em JHEP}
  {\bfseries 02} (2015) 172},
\href{http://arxiv.org/abs/1412.5148}{{\ttfamily arXiv:1412.5148 [hep-th]}}.

\bibitem{ThKe}
R.~Thorngren and C.~von Keyserlingk, ``{Higher SPT's and a generalization of
  anomaly in-flow},''
\href{http://arxiv.org/abs/1511.02929}{{\ttfamily arXiv:1511.02929
  [cond-mat.str-el]}}.

\bibitem{GKKS}
D.~Gaiotto, A.~Kapustin, Z.~Komargodski, and N.~Seiberg, ``{Theta, Time
  Reversal, and Temperature},''
  \href{http://dx.doi.org/10.1007/JHEP05(2017)091}{{\em JHEP} {\bfseries 05}
  (2017) 091},
\href{http://arxiv.org/abs/1703.00501}{{\ttfamily arXiv:1703.00501 [hep-th]}}.

\bibitem{TaKi}
Y.~Tanizaki and Y.~Kikuchi, ``{Vacuum structure of bifundamental gauge theories
  at finite topological angles},''
  \href{http://dx.doi.org/10.1007/JHEP06(2017)102}{{\em JHEP} {\bfseries 06}
  (2017) 102},
\href{http://arxiv.org/abs/1705.01949}{{\ttfamily arXiv:1705.01949 [hep-th]}}.

\bibitem{ShYo}
H.~Shimizu and K.~Yonekura, ``{Anomaly constraints on deconfinement and chiral
  phase transition},''
\href{http://arxiv.org/abs/1706.06104}{{\ttfamily arXiv:1706.06104 [hep-th]}}.

\bibitem{FuOs}
S.~C. Furuya and M.~Oshikawa, ``{Symmetry Protection of Critical Phases and a
  Global Anomaly in $1+1$ Dimensions},''
  \href{http://dx.doi.org/10.1103/PhysRevLett.118.021601}{{\em Phys. Rev.
  Lett.} {\bfseries 118} no.~2, (2017) 021601},
\href{http://arxiv.org/abs/1503.07292}{{\ttfamily arXiv:1503.07292
  [cond-mat.stat-mech]}}.

\bibitem{SeWi}
N.~Seiberg and E.~Witten, ``{Gapped Boundary Phases of Topological Insulators
  via Weak Coupling},'' \href{http://dx.doi.org/10.1093/ptep/ptw083}{{\em PTEP}
  {\bfseries 2016} no.~12, (2016) 12C101},
\href{http://arxiv.org/abs/1602.04251}{{\ttfamily arXiv:1602.04251
  [cond-mat.str-el]}}.

\bibitem{TaYo1}
Y.~Tachikawa and K.~Yonekura, ``{On time-reversal anomaly of 2+1d topological
  phases},'' \href{http://dx.doi.org/10.1093/ptep/ptx010}{{\em PTEP} {\bfseries
  2017} no.~3, (2017) 033B04},
\href{http://arxiv.org/abs/1610.07010}{{\ttfamily arXiv:1610.07010 [hep-th]}}.

\bibitem{TaYo2}
Y.~Tachikawa and K.~Yonekura, ``{More on time-reversal anomaly of 2+1d
  topological phases},''
  \href{http://dx.doi.org/10.1103/PhysRevLett.119.111603}{{\em Phys. Rev.
  Lett.} {\bfseries 119} no.~11, (2017) 111603},
\href{http://arxiv.org/abs/1611.01601}{{\ttfamily arXiv:1611.01601 [hep-th]}}.

\bibitem{Cardy}
J.~L. Cardy, ``{Boundary Conditions, Fusion Rules and the Verlinde Formula},''
\href{http://dx.doi.org/10.1016/0550-3213(89)90521-X}{{\em Nucl. Phys.}
  {\bfseries B324} (1989) 581--596}.

\bibitem{AKLT1}
I.~Affleck, T.~Kennedy, E.~H. Lieb, and H.~Tasaki, ``Rigorous results on
  valence-bond ground states in antiferromagnets,''
  \href{http://dx.doi.org/10.1103/PhysRevLett.59.799}{{\em Phys. Rev. Lett.}
  {\bfseries 59} (Aug, 1987) 799--802}.
  \url{https://link.aps.org/doi/10.1103/PhysRevLett.59.799}.

\bibitem{AKLT2}
I.~Affleck, T.~Kennedy, E.~H. Lieb, and H.~Tasaki, ``Valence bond ground states
  in isotropic quantum antiferromagnets,'' {\em Comm. Math. Phys.} {\bfseries
  115} no.~3, (1988) 477--528.
  \url{https://projecteuclid.org:443/euclid.cmp/1104161001}.

\bibitem{Kennedy}
T.~Kennedy, ``Exact diagonalisations of open spin-1 chains,'' {\em Journal of
  Physics: Condensed Matter} {\bfseries 2} no.~26, (1990) 5737.
  \url{http://stacks.iop.org/0953-8984/2/i=26/a=010}.

\bibitem{HTHR}
B.~Han, A.~Tiwari, C.-T. Hsieh, and S.~Ryu, ``{Boundary conformal field theory
  and symmetry protected topological phases in $2+1$ dimensions},''
  \href{http://dx.doi.org/10.1103/PhysRevB.96.125105}{{\em Phys. Rev.}
  {\bfseries B96} no.~12, (2017) 125105},
\href{http://arxiv.org/abs/1704.01193}{{\ttfamily arXiv:1704.01193
  [cond-mat.str-el]}}.

\bibitem{Bultinck:2017iff}
N.~Bultinck, R.~Vanhove, J.~Haegeman, and F.~Verstraete, ``{Global anomaly
  detection in two-dimensional symmetry-protected topological phases},''
\href{http://arxiv.org/abs/1710.02314}{{\ttfamily arXiv:1710.02314
  [cond-mat.str-el]}}.

\bibitem{FrVa}
D.~S. Freed and C.~Vafa, ``{GLOBAL ANOMALIES ON ORBIFOLDS},''
  \href{http://dx.doi.org/10.1007/BF01212418}{{\em Commun. Math. Phys.}
  {\bfseries 110} (1987) 349}.
[Addendum: Commun. Math. Phys.117,349(1988)].

\bibitem{FGK}
G.~Felder, K.~Gaw^^c4^^99dzki, and A.~Kupiainen, ``Spectra of
  wess-zumino-witten models with arbitrary simple groups,'' {\em Comm. Math.
  Phys.} {\bfseries 117} no.~1, (1988) 127--158.
  \url{https://projecteuclid.org:443/euclid.cmp/1104161597}.

\bibitem{SCR}
O.~M. Sule, X.~Chen, and S.~Ryu, ``{Symmetry-protected topological phases and
  orbifolds: Generalized Laughlin's argument},''
  \href{http://dx.doi.org/10.1103/PhysRevB.88.075125}{{\em Phys. Rev.}
  {\bfseries B88} (2013) 075125},
\href{http://arxiv.org/abs/1305.0700}{{\ttfamily arXiv:1305.0700
  [cond-mat.str-el]}}.

\bibitem{Vafa1986}
C.~Vafa, ``{Modular Invariance and Discrete Torsion on Orbifolds},''
\href{http://dx.doi.org/10.1016/0550-3213(86)90379-2}{{\em Nucl. Phys.}
  {\bfseries B273} (1986) 592--606}.

\bibitem{BCR}
M.~Billo, B.~Craps, and F.~Roose, ``{Orbifold boundary states from Cardy's
  condition},'' \href{http://dx.doi.org/10.1088/1126-6708/2001/01/038}{{\em
  JHEP} {\bfseries 01} (2001) 038},
\href{http://arxiv.org/abs/hep-th/0011060}{{\ttfamily arXiv:hep-th/0011060
  [hep-th]}}.

\bibitem{GeWi}
D.~Gepner and E.~Witten, ``{String Theory on Group Manifolds},''
\href{http://dx.doi.org/10.1016/0550-3213(86)90051-9}{{\em Nucl. Phys.}
  {\bfseries B278} (1986) 493--549}.

\bibitem{Yellow}
P.~Di~Francesco, P.~Mathieu, and D.~Senechal,
  \href{http://dx.doi.org/10.1007/978-1-4612-2256-9}{{\em {Conformal Field
  Theory}}}.
\newblock Graduate Texts in Contemporary Physics. Springer-Verlag, New York,
1997.
\newblock

\bibitem{DiWi}
R.~{Dijkgraaf} and E.~{Witten}, ``{Topological gauge theories and group
  cohomology},'' \href{http://dx.doi.org/10.1007/BF02096988}{{\em
  Communications in Mathematical Physics} {\bfseries 129} (Apr., 1990)
  393--429}.

\bibitem{AhWa}
C.-r. Ahn and M.~A. Walton, ``{Spectra of Strings on Nonsimply Connected Group
  Manifolds},''
\href{http://dx.doi.org/10.1016/0370-2693(89)91613-4}{{\em Phys. Lett.}
  {\bfseries B223} (1989) 343--348}.

\bibitem{GaSc}
B.~Gato-Rivera and A.~N. Schellekens, ``{Complete classification of simple
  current modular invariants for (Z(p))**k},''
\href{http://dx.doi.org/10.1007/BF02099282}{{\em Commun. Math. Phys.}
  {\bfseries 145} (1992) 85--122}.

\bibitem{Gaberdiel}
M.~R. Gaberdiel, ``{WZW models of general simple groups},''
  \href{http://dx.doi.org/10.1016/0550-3213(95)00587-0}{{\em Nucl. Phys.}
  {\bfseries B460} (1996) 181--202},
\href{http://arxiv.org/abs/hep-th/9508105}{{\ttfamily arXiv:hep-th/9508105
  [hep-th]}}.

\bibitem{Ishibashi}
N.~Ishibashi, ``{The Boundary and Crosscap States in Conformal Field
  Theories},''
\href{http://dx.doi.org/10.1142/S0217732389000320}{{\em Mod. Phys. Lett.}
  {\bfseries A4} (1989) 251}.

\bibitem{Onogi}
T.~Onogi and N.~Ishibashi, ``{Conformal Field Theories on Surfaces With
  Boundaries and Crosscaps},''
  \href{http://dx.doi.org/10.1142/S0217732389000228}{{\em Mod. Phys. Lett.}
  {\bfseries A4} (1989) 161}.
[Erratum: Mod. Phys. Lett.A4,885(1989)].

\bibitem{MMS}
J.~M. Maldacena, G.~W. Moore, and N.~Seiberg, ``{D-brane instantons and K
  theory charges},''
  \href{http://dx.doi.org/10.1088/1126-6708/2001/11/062}{{\em JHEP} {\bfseries
  11} (2001) 062},
\href{http://arxiv.org/abs/hep-th/0108100}{{\ttfamily arXiv:hep-th/0108100
  [hep-th]}}.

\bibitem{McGreevy}
J.~McGreevy, \href{http://dx.doi.org/10.1142/9789813149441_0004}{``{TASI 2015
  Lectures on Quantum Matter (with a View Toward Holographic Duality)},''} in
  {\em {Proceedings, Theoretical Advanced Study Institute in Elementary
  Particle Physics: New Frontiers in Fields and Strings (TASI 2015): Boulder,
  CO, USA, June 1-26, 2015}}, pp.~215--296.
\newblock 2017.
\newblock
\href{http://arxiv.org/abs/1606.08953}{{\ttfamily arXiv:1606.08953 [hep-th]}}.
\newblock

\bibitem{Duivenvoorden:2012yr}
K.~Duivenvoorden and T.~Quella, ``{Topological phases of spin chains},''
  \href{http://dx.doi.org/10.1103/PhysRevB.87.125145}{{\em Phys. Rev.}
  {\bfseries B87} no.~12, (2013) 125145},
\href{http://arxiv.org/abs/1206.2462}{{\ttfamily arXiv:1206.2462
  [cond-mat.str-el]}}.

\bibitem{Roy:2015ars}
A.~Roy and T.~Quella, ``{Chiral Haldane phases of SU(N) quantum spin chains in
  the adjoint representation},''
\href{http://arxiv.org/abs/1512.05229}{{\ttfamily arXiv:1512.05229
  [cond-mat.str-el]}}.

\bibitem{PTBO}
F.~Pollmann, A.~M. Turner, E.~Berg, and M.~Oshikawa, ``Entanglement spectrum of
  a topological phase in one dimension,''
  \href{http://dx.doi.org/10.1103/PhysRevB.81.064439}{{\em Phys. Rev. B}
  {\bfseries 81} (Feb, 2010) 064439}.
  \url{https://link.aps.org/doi/10.1103/PhysRevB.81.064439}.

\bibitem{TaTo}
K.~{Tanimoto} and K.~{Totsuka}, ``{Symmetry-protected topological order in
  SU(N) Heisenberg magnets --quantum entanglement and non-local order
  parameters},'' {\em ArXiv e-prints} (Aug., 2015) ,
  \href{http://arxiv.org/abs/1508.07601}{{\ttfamily arXiv:1508.07601
  [cond-mat.str-el]}}.

\bibitem{CaTo}
J.~Cardy and E.~Tonni, ``{Entanglement hamiltonians in two-dimensional
  conformal field theory},''
  \href{http://dx.doi.org/10.1088/1742-5468/2016/12/123103}{{\em J. Stat.
  Mech.} {\bfseries 1612} no.~12, (2016) 123103},
\href{http://arxiv.org/abs/1608.01283}{{\ttfamily arXiv:1608.01283
  [cond-mat.stat-mech]}}.

\bibitem{ACT}
V.~Alba, P.~Calabrese, and E.~Tonni, ``{Entanglement spectrum degeneracy and
  the Cardy formula in 1+1 dimensional conformal field theories},''
  \href{http://dx.doi.org/10.1088/1751-8121/aa9365}{{\em J. Phys.} {\bfseries
  A51} no.~2, (2018) 024001},
\href{http://arxiv.org/abs/1707.07532}{{\ttfamily arXiv:1707.07532 [hep-th]}}.

\bibitem{BrWi}
J.~C. Bridgeman and D.~J. Williamson, ``{Anomalies and entanglement
  renormalization},'' \href{http://dx.doi.org/10.1103/PhysRevB.96.125104}{{\em
  Phys. Rev.} {\bfseries B96} no.~12, (2017) 125104},
\href{http://arxiv.org/abs/1703.07782}{{\ttfamily arXiv:1703.07782
  [quant-ph]}}.

\bibitem{SiVi}
S.~Singh and G.~Vidal, ``{Symmetry protected entanglement renormalization},''
  \href{http://dx.doi.org/10.1103/PhysRevB.88.121108}{{\em Phys. Rev.}
  {\bfseries B88} no.~12, (2013) 121108},
\href{http://arxiv.org/abs/1303.6716}{{\ttfamily arXiv:1303.6716
  [cond-mat.str-el]}}.

\bibitem{EveVi}
G.~{Evenbly} and G.~{Vidal}, ``{Algorithms for Entanglement Renormalization:
  Boundaries, Impurities and Interfaces},''
  \href{http://dx.doi.org/10.1007/s10955-014-0983-1}{{\em Journal of
  Statistical Physics} (Apr., 2014) },
  \href{http://arxiv.org/abs/1312.0303}{{\ttfamily arXiv:1312.0303
  [quant-ph]}}.

\end{thebibliography}\endgroup

\end{document}